\documentclass[twocolumn,english,aps,prb]{revtex4-1}
\usepackage[T1]{fontenc}
\usepackage[latin9]{inputenc}
\usepackage{color}
\usepackage{amsmath}
\usepackage{amssymb}
\usepackage{graphicx}

\makeatletter
\@ifundefined{textcolor}{}
{%
 \definecolor{BLACK}{gray}{0}
 \definecolor{WHITE}{gray}{1}
 \definecolor{RED}{rgb}{1,0,0}
 \definecolor{GREEN}{rgb}{0,1,0}
 \definecolor{BLUE}{rgb}{0,0,1}
 \definecolor{CYAN}{cmyk}{1,0,0,0}
 \definecolor{MAGENTA}{cmyk}{0,1,0,0}
 \definecolor{YELLOW}{cmyk}{0,0,1,0}
}

\usepackage{babel}

\makeatother

\usepackage{babel}
\begin{document}

\title{The effect of magnetic field on the intrinsic detection efficiency
of superconducting single-photon detectors}

\author{J.J. Renema}

\author{R.J. Rengelink}

\author{I. Komen}

\author{Q. Wang}

\affiliation{Huygens-Kamerlingh Onnes lab, Leiden University, Niels Bohrweg 2,
2333 CA Leiden, the Netherlands}

\author{R. Gaudio}

\author{K.P.M. op 't Hoog}

\author{Z. Zhou}

\affiliation{COBRA Research Institute, Eindhoven University of Technology, P.O.
Box 513, 5600 MB Eindhoven, The Netherlands}

\author{D. Sahin}

\affiliation{COBRA Research Institute, Eindhoven University of Technology, P.O.
Box 513, 5600 MB Eindhoven, The Netherlands}

\affiliation{Centre for Quantum Photonics, H. H. Wills Physics Laboratory, University
of Bristol, Tyndall Avenue, Bristol BS8 1TL, the UK}

\selectlanguage{english}%

\author{A. Fiore}

\affiliation{COBRA Research Institute, Eindhoven University of Technology, P.O.
Box 513, 5600 MB Eindhoven, The Netherlands}

\author{P. Kes}

\author{J. Aarts}

\author{M.P. van Exter}

\author{M.J.A. de Dood}

\affiliation{Huygens-Kamerlingh Onnes lab, Leiden University, Niels Bohrweg 2,
2333 CA Leiden, the Netherlands}

\author{E.F.C. Driessen}

\affiliation{Univ. Grenoble Alpes, INAC-SPSMS, 38000 Grenoble, France}

\affiliation{CEA, INAC-SPSMS, 38000 Grenoble, France}
\begin{abstract}
We experimentally investigate the effect of a magnetic field on photon
detection in superconducting single-photon detectors. At low fields,
the effect of a magnetic field is through the direct modification
of the quasiparticle density of states of the superconductor, and
magnetic field and bias current are interchangable, as is expected
for homogeneous dirty-limit superconductors. At the field where a
first vortex enters the detector, the effect of the magnetic field
is reduced, up until the point where the critical current of the detector
starts to be determined by flux flow. From this field on, increasing
the magnetic field does not alter the detection of photons anymore,
whereas it does still change the rate of dark counts. This result
points at an intrinsic difference in dark and light counts, and also
shows that no enhancement of the intrinsic detection efficiency of
a straight SSPD wire is achievable in a magnetic field.
\end{abstract}
\maketitle
Nanowire superconducting single-photon detectors \cite{Goltsman2001}
are a crucial technology for single-photon detection in the infrared,
since they can achieve detection efficiencies of up to 93\% \cite{Marsili},
with low dark count rate, low jitter, and short reset time \cite{Natarajanrevie}.
These detectors consist of a\textcolor{black}{{} narrow and thin wire}
of superconducting material, carrying a bias current. 

\textcolor{black}{While the broad outlines of the photodetection mechanism
are known, there is as yet no complete theory describing the response
of such detectors. The present understanding of photodetection in
SSPDs is as follows }\cite{Engelpreprint,RenemaPRB,RenemaPRL,ZotovaARXIV,VodolazovPRB,Eftekhariangroundstates,Bulaevskii2012,Bulaevskii2011,Gurevich2012}\textcolor{black}{:
when a photon is absorbed, a cloud of quasiparticles is created which
locally reduces the current-carrying capacity of the wire. }Current
is expelled from the absorption spot. If this diverted current is
sufficiently strong, which depends on both the initial bias current
and the energy of the photon, the Lorentz force may cause the unbinding
of a vortex from the edge of the wire, \textcolor{black}{leading}
to a measurable voltage pulse. Therefore, experiments on SSPDs in
a magnetic field are a natural way of investigating the detection
mechanism; \textcolor{black}{o}ne might even wonder whether the efficiency
of the detector could be enhanced by applying a magnetic field. 

In the present work, we study how an applied magnetic field directly
affects the microscopic detection mechanism in a short section of
wire. By using a single narrow active area in a bridge-like configuration,
we avoid the question of current flow around curved sections of the
device, which complicated the interpretation of previous experiments
\cite{Luschepreprint,LuscheIEEE,Engelmagneticfields}. We find that
it is the direct modification of the quasiparticle density of states
in the superconductor that governs the magnetic field behaviour of
SSPDs. In dirty-limit superconductors (such as thin-film NbN), this
density of states is modified by a bias current or a magnetic field
\cite{Anthore}. The effect of a magnetic field is therefore a homogeneous
weakening of Cooper pairing, resulting in a higher detection efficiency
at constant bias current. We identify three regimes. In the low-field
regime (up to $\sim$ 50 mT) the current flow is sufficiently homogeneous.
Bias current and magnetic field are completely interchangeable, as
described by the Usadel equations \cite{Usadel}. The relevant parameters
of this theory do not depend on the illumination wavelength or on
temperature in our measurement range, as is expected. In the intermediate
regime (50 mT - 200 mT) we still observe light counts, but a higher
current is required to achieve photodetection than predicted by the
homogeneous theory. In the high-field regime (200 mT), first light
and then dark counts are gradually extinguished when the field is
increased. We attribute this to the presence of vortices in the wire.

We find that the enhancement of light and dark counts on a single
active spot obey different field scales, pointing to a fundamental
difference in the nature of the two. The field scale for the reduction
of the critical current is smaller than the scale for the increase
of the count rate. This leads us to conclude that no intrinsic enhancement
of the detection efficiency of an SSPD under the influence of a magnetic
field is possible. 

\textcolor{black}{Our experiments were performed on two different
detectors: a 200~nm long bridge with a width of 150~nm (sample A),
and a bowtie-shaped nanodetector \cite{Bitauld2010} with a width
of 220~nm (sample B). The detectors were fabricated on 5~nm th}ick
NbN films, that were sputter-deposited on a GaAs substrate. The detectors
were patterned using conventional e-beam lithography and reactive-ion
etching in a \textcolor{black}{$\mathrm{SF}{}_{6}$/ Ar plasma \cite{Gaggero2010}.
}After patterning, detectors had a critical temperature \textcolor{black}{of
9.5~K, }and a sheet resistance of $R_{\square}=600\ \Omega$. 

\textcolor{black}{The samples were mounted in a Physical Properties
Measurement System (PPMS)}\textcolor{blue}{{} }in a custom insert that
allows optical coupling and high-frequency electronic readout. %
\footnote{We verified that the temperature of this custom insert was identical
to the temperature measured on the block thermometer of the PPMS.%
} \textcolor{black}{We bias our device through a 100 $\Omega$ resistor,
and measure the current and voltage over the device. We determine
the critical current with a 10 $\Omega$ resistance }criterion. The
noise from the room-temperature broadband amplifiers in our measurement
circuit reduces the critical current of our devices by approximately
1 $\mu$A. To facilitate comparison between critical-current measurements
and count rate measurements, all measurements presented here were
performed with these amplifiers present in the circuit. 

\textcolor{black}{The orientation of the applied magnetic field was
perpendicular to the film. In order to avoid hysteresis, all measurements
were performed while increasing the magnetic field. After a measurement
run, the field was further increased to 1~T, before ramping it down
using the demagnetizing (degaussing) option of the PPMS control software},
which was found to be crucial for obtaining reproducible results.
The remanent field was estimated to be 1 mT, which is consistent with
specifications \cite{QDmanual}. 

\textcolor{black}{We illuminate our detectors with a continuous-wave
laser with a wavelength of $826$ nm, and an optical power of }12
mW. The illumination spot is approximately 2 mm in diameter. We have
no control over the polarization, but it was kept constant during
the experiment. \textcolor{black}{We recorded the count rate during
a 1 s interval at each current. }

\textcolor{black}{In Fig. 1, we plot a typical experimental result.
The magnetic field was increased from 0 mT to 300 mT in steps of 30
mT. We observe an exponential increase of the count rate with bias
current, followed by a saturation at higher currents and a final exponential
increase associated with dark counts, as is usually observed for this
kind of detectors \cite{Natarajan2010}. The presence of a magnetic
field shifts the curve towards lower currents }%
\footnote{\textcolor{black}{From the fact that the curves have the typical shape
for 1-photon detection and from the low count rate, we infer that
multiphoton detection events do not play a role. }%
}\textcolor{black}{. We note that as the field is increased, a larger
part of the count rate curve is dominated by dark counts. We conclude
that light counts and dark counts obey different field scales, even
in our geometry where there is a single active area.}

\textcolor{black}{We have compared our results to the theory of Bulaevskii
}\textcolor{black}{\emph{et al}}\textcolor{black}{.\cite{Bulaevskii2011,Bulaevskii2012},
which considers the effect of a magnetic field on the entry barrier
of vortices. This theory predicts an exponential increase of count
rate as a function of applied field, at constant bias current. As
in previous experiments \cite{Engelmagneticfields,Luschefields},
we find that prediction this theory gives for the rate of exponential
increase is an order of magnitude away from the experimental value.}

\begin{figure}
\includegraphics[width=87mm]{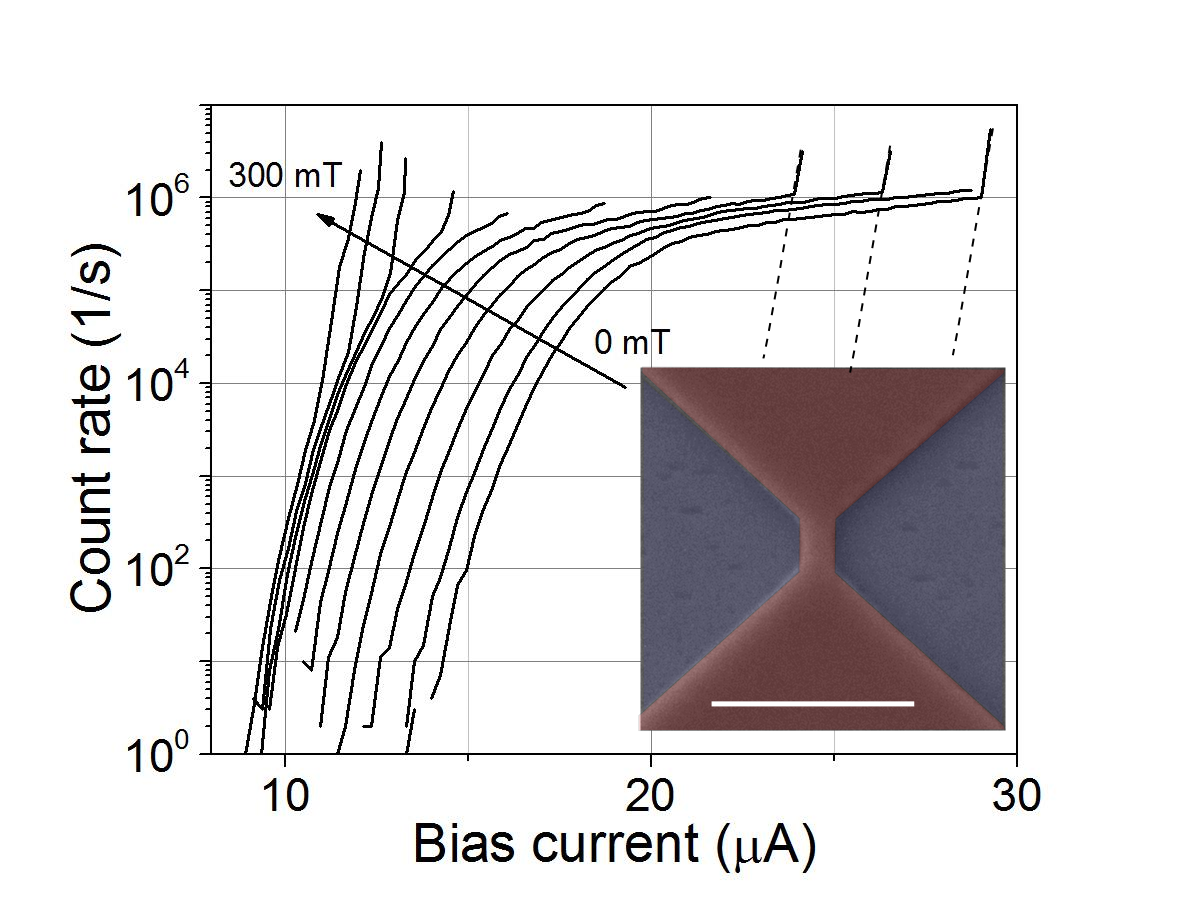}\protect\caption{Count rate of sample A, illuminated with 826 nm light at T = 1.8 K,
for different magnetic fields ranging from 0 mT to 300 mT, in steps
of 30 mT. This measurement was not corrected for dark counts\emph{.
}The dashed lines are a guide to the eye indicating the part of the
curve where dark counts are dominant.\emph{ Inset} false-colour SEM
image of a detector (NbN coloured red) nominally identical to sample
A. The scale bar is 1 $\mu\mathrm{m}.$ }
\end{figure}

In figure 2, we plot those combinations of bias current and magnetic
field which are required to achieve a constant count rate, from $1$/s
to $10^{5}$/s. For low magnetic fields $B\lesssim50$ mT, the resulting
iso-count rate curves lie on a series of concentric ellipses, which
we have plotted in figure 2. For sample B, we similarly find concentric
elipses (not shown). In the measurement regime reported here, the
dark count rate is negligible ($\ll$ 1 / s).

\begin{figure}
\includegraphics[width=87mm]{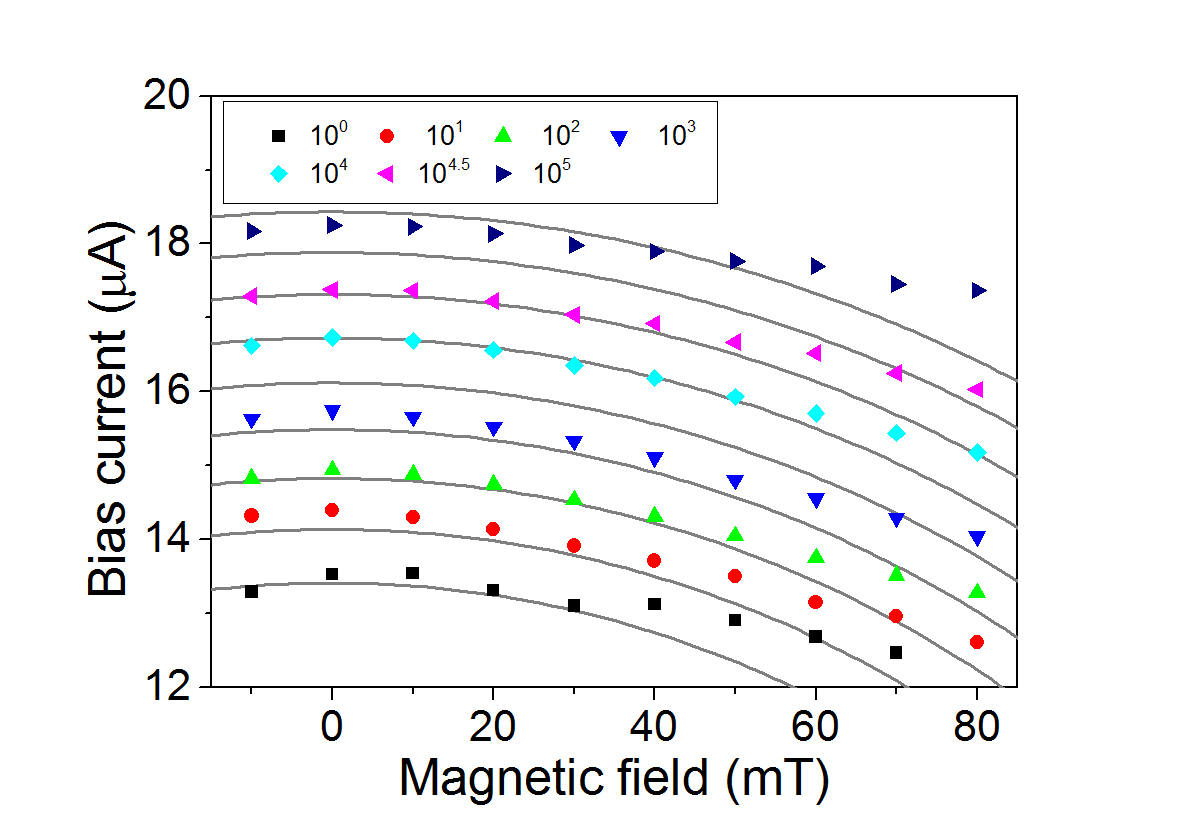}

\protect\caption{Bias current at constant count rate as a function of magnetic field
for sample A. The different colours and symbols correspond to different
count rates, over five orders of magnitude. We find that for low magnetic
fields (up to $\sim$ 50 mT) the required current to achieve a certain
number of counts depends quadratically on applied magnetic field.
The grey lines are equidistant elipses calculated using the Usadel
formalism (see text). }
\end{figure}

In Fig. 3, we turn to the temperature dependence of the magnetic-field
response. We find that changing the temperature induces an overall
shift in the iso-count-rate curves, but that $I_{\Gamma}$ and $B_{\Gamma}$
are independent of temperature. We have also verified these parameters
are independent of illumination wavelength by repeating the experiment
with light of 405 nm and 1300 nm (not shown). The shift in count rate
as a function of temperature at zero field is consistent with our
previous results \cite{RenemaPRL}, where we showed that the temperature
dependence of the SSPD response is determined by the energy barrier
for vortex entry. 

\begin{figure}
\includegraphics[width=87mm]{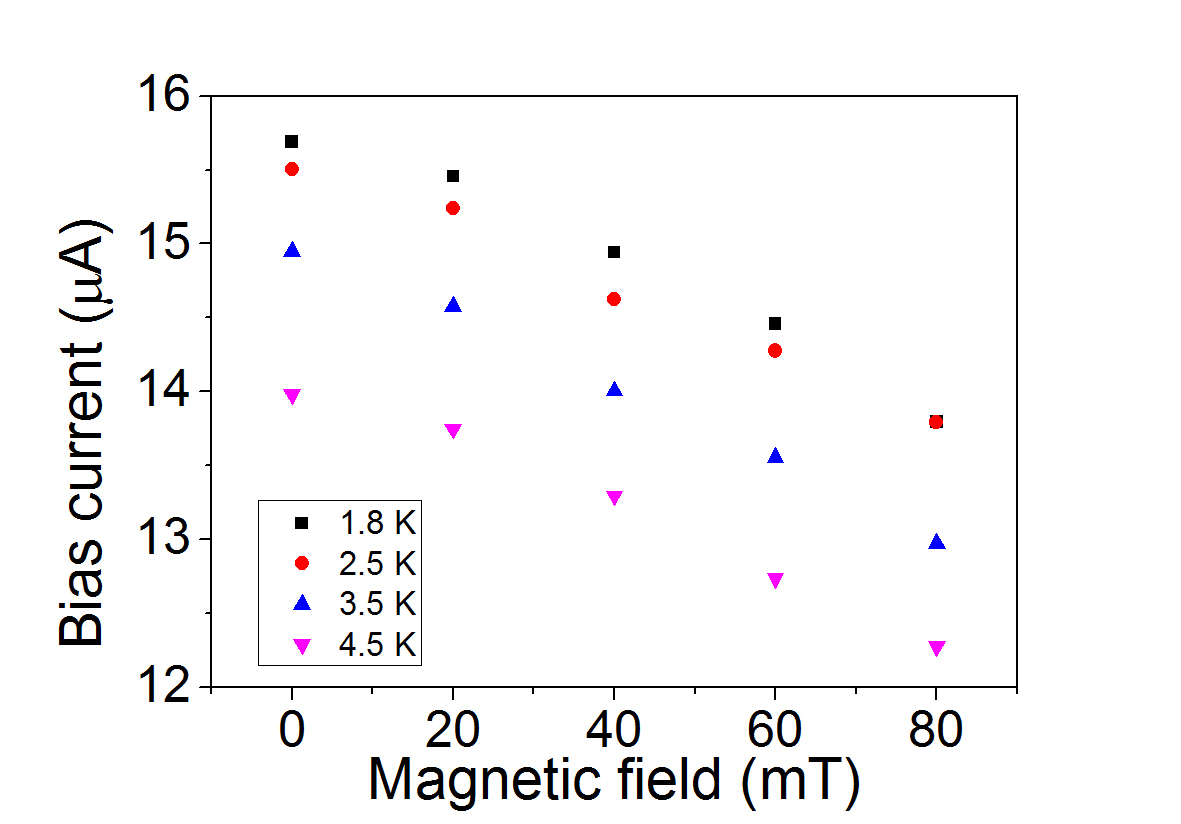}

\protect\caption{Magnetic field dependence of the count rate at different temperatures
for sample A. We plot the count rate required to obtain 1000 counts
/ s at different temperatures. We find that $I_{\Gamma}$ and $B_{\Gamma}$
are independent of temperature. }
\end{figure}

\textcolor{black}{We analyze these observations in terms of the microscopic
theory for dirty-limit superconductivity, motivated by our analysis
of the modification of the electronic state due to intrinsic pair
breaking in similar films\cite{DriessenPRL2012}. For our film, $\lambda_{\perp}\gg w,d$,
with $\lambda_{\perp}$ the effective penetration length, and $w$
and $d$ the width and thickness of the wire, respectively. Therefore,
we assume a homogeneous current flow through our wire. In this case,
the superconducting state is described by the homogeneous Usadel equation
\cite{Usadel}: }

\begin{equation}
iE\sin\theta+\Delta\cos\theta-\Gamma\sin\theta\cos\theta=0,
\end{equation}
\textcolor{black}{where $E$ is the quasiparticle energy, $\theta$
is the pairing angle, $\Delta$ the superconducting pairing potential
and $\Gamma$ the pair breaking energy, representing a finite momentum
of the Cooper pairs. A bias current $I_{b}$ and a perpendicular magnetic
field have a similar effect in weakening the superconducting state,
as was shown experimentally by Anthore }\textcolor{black}{\emph{et
al.}}\textcolor{black}{{} for one-dimensional aluminium wires\cite{Anthore}.
In this case, the depairing energy can be approximated by: 
\begin{equation}
\frac{\Gamma}{\Delta}=\left(\frac{\Delta}{U(\Gamma)}\frac{I_{b}}{I_{\Gamma}}\right)^{2}+\left(\frac{B}{B_{\Gamma}}\right)^{2},
\end{equation}
\begin{equation}
\frac{U(\Gamma)}{\Delta}\approx\frac{\pi}{2}-1.8\frac{\Gamma}{\Delta}-\left(\frac{\Gamma}{\Delta}\right)^{2},
\end{equation}
where $I_{\Gamma}=\sqrt{2}\Delta/eR(\xi)$ and $B_{\Gamma}=\sqrt{6}(\hbar/ew\xi)$
are characteristic current and field scales, respectively, with $R(\xi)$
the resistance of a section of the wire one coherence length $\xi$
long.}

\textcolor{black}{We note that the structure of these equations is
compatible with our experimental observations at low fields: they
define a series of concentric ellipses in the $I-B$ plane, connecting
points with equal value of $\Gamma/\Delta$. For a more quantitative
analysis of $I_{\Gamma}$ and $B_{\Gamma}$, we have determined the
coherence length $\xi$=3.9~nm from the slope of the upper critical
field at the critical temperature. To evaluate}\textcolor{red}{{} }\textcolor{black}{$R(\xi)$
=}\textcolor{blue}{{} }\textcolor{black}{7.2 $\Omega$, we have assumed
a homogeneous sheet resistance of our NbN film. }We have determined
the value of $\Delta$ = 1.9 meV at T = 1.5 K using scanning tunnelling
spectroscopy on a piece of the same film that was used to fabricate
the detectors. \textcolor{black}{In the STM tunneling spectra, we
observe slightly rounded-off coherence peaks, consistent with the
presence of an intrinsic pair breaker $\Gamma\approx100\:\mu eV$,
as was found previously on NbTiN and TiN films with similar resistivity
\cite{DriessenPRL2012,CoumouPRB}. The presence of this pair breaker
does not change the analysis that we present here. }Using these values,
we estimate \textcolor{black}{$I_{\Gamma}=180\pm20\:\mu$A, }$B_{\Gamma}=2.7$
T for sample A, and $I_{\Gamma}=330\pm20$ $\mu$A, $B_{\Gamma}=1.8$
T for sample B. These values were used in generating the elipses in
figure 2; the only remaining freedom is the dependence between the
count rate $C$ and the normalized pair breaking energy $C(\Gamma/\Delta)$. 

From the excellent agreement beween theory and experiment at magnetic
field values B$\lesssim50~\mathrm{mT}$, we conclude that in this
regime the count rate of the detector is determined only by a homogeneous
weakening of the superconducting state, that can be described by the
depairing energy $\Gamma$. This implies that the only way in which
the magnetic field affects the detection mechanism is through the
electronic state of the superconducting film before a vortex is absorbed.
This picture is reenforced by the fact that the effects of magnetic
field and temperature are independent: the field response is set by
the film, whereas the temperature response is set by the barrier for
a vortex entering the wire when a detection event occurs.

\begin{figure}
\includegraphics[width=87mm]{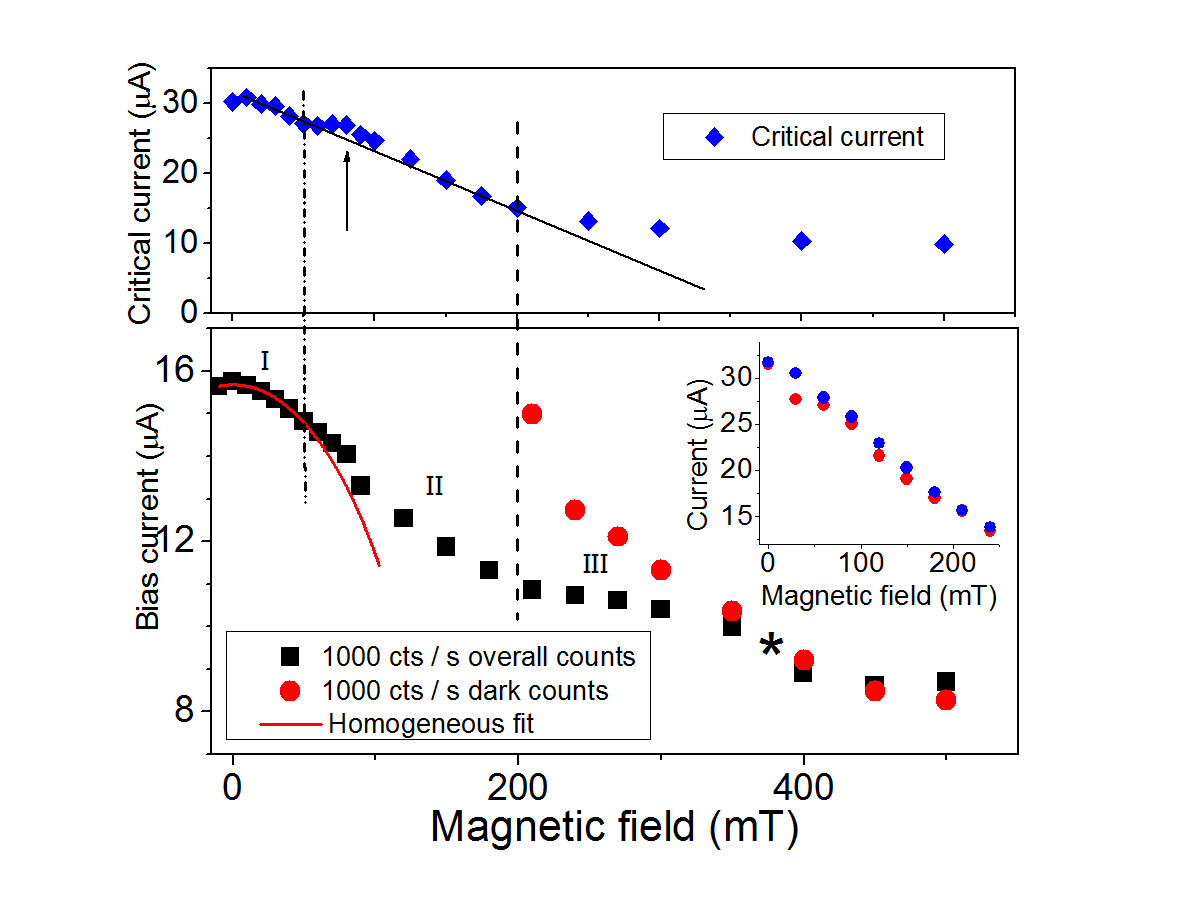}\protect\caption{Critical current (\emph{top}) and curve of constant count rate (\emph{bottom})
as a function of magnetic field for sample A. The black squares indicate
count rate under illumination (light counts + dark counts), the red
circles indicate dark counts and the blue diamonds indicate critical
current. The asterisk marks the point where all observed counts are
dark counts. The solid line in the top panel is a guide to the eye.
The red curve a plot of Eq. (2) for this count rate. We identify three
regimes (see text), demarcated by the two vertical lines. \emph{Inset:
}Critical current and 1000/s dark count rate at low fields. }
\end{figure}

In Fig. 4, we plot the field dependence of a representative count
rate (1000/s) and the field dependence of the critical current for
a wider range of magnetic fields. We phenomenologically distinguish
three regimes, independent of the chosen count rate. In the first
regime, up to B $\lesssim$ 50 mT, our data follows the prediction
from the homogeneous theory. In the second regime ($50~\mathrm{mT}<B<200~\mathrm{mT}$),
more current is required to produce detection events with a given
probability than predicted by the homogeneous theory. In the third
regime ($B>200~\mathrm{mT})$, the count rate is almost independent
of the applied field. However, the critical current continues to decrease
and we observe throughout our measurement range that the dark counts
shift with the critical current (see inset fig 3). Eventually, there
is a count-rate dependent point where the observed count rate is entirely
comprised of dark counts, indicated in Fig 4. with an asterisk. At
a magnetic field of approximately 1 T, no detection events are observed
anymore in a 1s interval.

\textcolor{black}{To understand the physical meaning of the three
regimes, we turn to the critical current measurements, shown in the
top panel of figure 4. We observe linear decay of the form $I_{c}(B)=I_{c}(0)(1-B/B_{0})$,
with $B_{0}=375$ mT, up to the point $I_{c}(B)=0.5I_{c}(0)$. At
higher fields, we obtain a power-law behavior $I_{c}\propto B^{\alpha}$,
with $\alpha\approx-0.4$. In this regime we observe that there is
no sharp transition to the normal state. We interpret these results
in the context of the extensive literature on the field dependence
of the critical current of superconducting strips, where the transition
from linear to power-law behaviour is interpreted as the transition
from a regime of critical current set by induced depairing to a regime
where the critical current is set by flux flow \cite{Beanlivingston,Kesedgebarrier2001,Ilinwidthjc,IlinphysicaC}.
T}he transition from induced depairing to flux flow corresponds to
the transition of regime II to regime III in Fig.~4.

\textcolor{black}{One important difference from previous results is
the additional feature indicated by an arrow in the critical current
measurements around 80 mT, where the critical current is enhanced
relative to the linear dependence. We interpret this feature as a
single vortex which is pinned in our material. All our measurements
were done in a geometry that is intrinsically photodetecting, and
a photodetection event entails a transition of the wire the normal
state and Joule heating. Therefore, in-field cooling occurs each time
there is a detection event. At 90 mT, we meet the criterion \cite{Stanmartinis}
for entry of the first vortex $B\approx\Phi_{0}/w^{2}$. We conclude
that while there is still an edge barrier at B = 80 mT, so that vortices
cannot enter, apparently the pinning is strong enough that a vortex
which is already there is not expelled. We note that Il'in }\textcolor{black}{\emph{et
al}}\textcolor{black}{{} \cite{IlinIC_PRB} have seen comparable enhancements
of the critical current that were due to vortices, albeit in the flux-flow
regime. }

\textcolor{black}{From this, we infer the following explanation of
our results: in regime I, the current flow is sufficiently homogeneous
so that the response can be explained by a homogeneous degradation
of the superconducting state, described by the homogeneous Usadel
equation. At the beginning of regime II, a vortex enters in the detector
and is pinned in the material. This destroys the homogeneity. From
the fact that the current which is required to obtain a detection
event is higher than expected from theory, we infer that the presence
of a vortex is detrimental to the detection process.}

A full theory of regimes II and III is beyond the scope of the present
work. It would have to take into account the direct effect of the
magnetic field on the vortex barrier, the current distribution in
our sample in the presence of vortices, and the associated local changes
in $\Delta$. Any full microscopic theory of photodetection in SSPDs,
even at zero magnetic field, would also need to take into account
the intrinsic inhomogeneity that has been observed in similar \textcolor{black}{films
\cite{SacepePRL,NoatPRB}}, and the observed intrinsic pair breaker,
as it has been shown recently that these can give rise to an unexpected
response to electromagnetic radiation \textcolor{black}{\cite{BuenoARXIV}}.

We have demonstrated that for low fields, the response of an SSPD
to an applied magnetic field is set entirely by the effect that the
field has on the electronic state of the material. In this regime,
there is an interchange between bias current and applied magnetic
field, in agreement with the homogeneous theory for dirty-limit superconductivity.
Since the material parameters that enter this theory ($\lambda_{\perp},\rho$)
are similar for all SSPDs found in literature, our results are not
limited to NbN detectors. WSi, for example has $\rho=200\ \mu\Omega\mathrm{cm}$
and $\lambda=1400$ nm \cite{Marsili,EngelarXiv2}. For the intermediate
and high-field regimes, geometry and flux pinning properties become
more relevant. Therefore a more diverse behaviour might be expected.

Our experiment disproves that the difference between light and dark
counts in a magnetic field is due to them originating from different
points in the wire, as has been put forward by others \cite{Berdiyorovturns2012,Zotovaturns2013,Luschefields}.
We conclude that there is a difference in the nature of light and
dark counts in SSPDs: light counts scale with a field scale $B_{\Gamma}$
inherent to the material, whereas dark counts scale with the change
in critical current under the influence of magnetic field, which depends
on geometry. This difference between light and dark counts is as of
yet unexplained and carries implications for the design of SSPDs:
it means that the only way in which an SSPD can be made more efficient
by an applied magnetic field is by choosing a geometry where the critical
current is not adversely affected by the applied field, such as a
spiral \cite{Henrichspiral}. For a straight wire, we conclude that
no enhancement of the detection efficiency can be achieved by applying
a magnetic field.
\begin{acknowledgments}
We would like to thank M. Hoek, M. Rosticher, R. Lusche, A. Engel
and C. Chapelier for discussions, J. Mesman for construction of the
insert, M. Hesselberth, D. Boltje, K. Uhlirova and I.M. Dildar for
assistance with the experimental apparatus, B. Krama, R. Koehler and
A. van Amersfoort for assistance with the electronic circuit, and
M. van Kralingen for his assistance in the early stages of this experiment.\textcolor{black}{{}
}This work is part of the research programme of the Foundation for
Fundamental Research on Matter (FOM), which is financially supported
by the Netherlands Organisation for Scientific Research (NWO). It
is also supported by NanoNextNL, a micro- and nanotechnology program
of the Dutch Ministry of Economic Affairs, Agriculture and Innovation
(EL\&I) and 130 partners,\textcolor{black}{{} and by the Dutch Technology
Foundation STW, applied science division of NWO, the Technology Program
of the Ministry of Economic Affairs under Project No. 10380. E.F.C.D.
was financially supported by the CEA-Eurotalents programme.}
\end{acknowledgments}


\begin{thebibliography}
\expandafter\ifx\csname natexlab\endcsname\relax\def\natexlab#1{#1}\fi
\expandafter\ifx\csname bibnamefont\endcsname\relax
  \def\bibnamefont#1{#1}\fi
\expandafter\ifx\csname bibfnamefont\endcsname\relax
  \def\bibfnamefont#1{#1}\fi
\expandafter\ifx\csname citenamefont\endcsname\relax
  \def\citenamefont#1{#1}\fi
\expandafter\ifx\csname url\endcsname\relax
  \def\url#1{\texttt{#1}}\fi
\expandafter\ifx\csname urlprefix\endcsname\relax\def\urlprefix{URL }\fi
\providecommand{\bibinfo}[2]{#2}
\providecommand{\eprint}[2][]{\url{#2}}

\bibitem[{\citenamefont{Goltsman et~al.}(2001)\citenamefont{Goltsman, Okunev,
  Chulkova, Lipatov, Semenov, Smirnov, Voronov, Dzardanov, Williams, and
  Sobolewski}}]{Goltsman2001}
\bibinfo{author}{\bibfnamefont{G.~N.} \bibnamefont{Goltsman}},
  \bibinfo{author}{\bibfnamefont{O.}~\bibnamefont{Okunev}},
  \bibinfo{author}{\bibfnamefont{G.}~\bibnamefont{Chulkova}},
  \bibinfo{author}{\bibfnamefont{A.}~\bibnamefont{Lipatov}},
  \bibinfo{author}{\bibfnamefont{A.}~\bibnamefont{Semenov}},
  \bibinfo{author}{\bibfnamefont{K.}~\bibnamefont{Smirnov}},
  \bibinfo{author}{\bibfnamefont{B.}~\bibnamefont{Voronov}},
  \bibinfo{author}{\bibfnamefont{A.}~\bibnamefont{Dzardanov}},
  \bibinfo{author}{\bibfnamefont{C.}~\bibnamefont{Williams}}, \bibnamefont{and}
  \bibinfo{author}{\bibfnamefont{R.}~\bibnamefont{Sobolewski}},
  \bibinfo{journal}{Appl. Phys. Lett.} \textbf{\bibinfo{volume}{79}},
  \bibinfo{pages}{705} (\bibinfo{year}{2001}).

\bibitem[{\citenamefont{Marsili et~al.}(2013)\citenamefont{Marsili, Verma,
  Stern, Harrington, Lita, Gerrits, Vayshenker, and Baek}}]{Marsili}
\bibinfo{author}{\bibfnamefont{F.}~\bibnamefont{Marsili}},
  \bibinfo{author}{\bibfnamefont{V.~B.} \bibnamefont{Verma}},
  \bibinfo{author}{\bibfnamefont{J.~A.} \bibnamefont{Stern}},
  \bibinfo{author}{\bibfnamefont{S.}~\bibnamefont{Harrington}},
  \bibinfo{author}{\bibfnamefont{A.~E.} \bibnamefont{Lita}},
  \bibinfo{author}{\bibfnamefont{T.}~\bibnamefont{Gerrits}},
  \bibinfo{author}{\bibfnamefont{I.}~\bibnamefont{Vayshenker}},
  \bibnamefont{and} \bibinfo{author}{\bibfnamefont{B.}~\bibnamefont{Baek}},
  \bibinfo{journal}{Nat. Photonics} \textbf{\bibinfo{volume}{7}},
  \bibinfo{pages}{210} (\bibinfo{year}{2013}).

\bibitem[{\citenamefont{Natarajan et~al.}(2012)\citenamefont{Natarajan, Tanner,
  and Hadfield}}]{Natarajanrevie}
\bibinfo{author}{\bibfnamefont{C.}~\bibnamefont{Natarajan}},
  \bibinfo{author}{\bibfnamefont{M.}~\bibnamefont{Tanner}}, \bibnamefont{and}
  \bibinfo{author}{\bibfnamefont{R.}~\bibnamefont{Hadfield}},
  \bibinfo{journal}{Superconductor Science and Technology}
  \textbf{\bibinfo{volume}{25}}, \bibinfo{pages}{063001}
  (\bibinfo{year}{2012}).

\bibitem[{\citenamefont{Engel and Schilling}(2013)}]{Engelpreprint}
\bibinfo{author}{\bibfnamefont{A.}~\bibnamefont{Engel}} \bibnamefont{and}
  \bibinfo{author}{\bibfnamefont{A.}~\bibnamefont{Schilling}},
  \bibinfo{journal}{J. Appl. Phys.} \textbf{\bibinfo{volume}{114}},
  \bibinfo{eid}{214501} (\bibinfo{year}{2013}).

\bibitem[{\citenamefont{Renema et~al.}(2013)\citenamefont{Renema, Frucci, Zhou,
  Mattioli, Gaggero, Leoni, de~Dood, Fiore, and van Exter}}]{RenemaPRB}
\bibinfo{author}{\bibfnamefont{J.~J.} \bibnamefont{Renema}},
  \bibinfo{author}{\bibfnamefont{G.}~\bibnamefont{Frucci}},
  \bibinfo{author}{\bibfnamefont{Z.}~\bibnamefont{Zhou}},
  \bibinfo{author}{\bibfnamefont{F.}~\bibnamefont{Mattioli}},
  \bibinfo{author}{\bibfnamefont{A.}~\bibnamefont{Gaggero}},
  \bibinfo{author}{\bibfnamefont{R.}~\bibnamefont{Leoni}},
  \bibinfo{author}{\bibfnamefont{M.~J.~A.} \bibnamefont{de~Dood}},
  \bibinfo{author}{\bibfnamefont{A.}~\bibnamefont{Fiore}}, \bibnamefont{and}
  \bibinfo{author}{\bibfnamefont{M.}~\bibnamefont{van Exter}},
  \bibinfo{journal}{Phys. Rev. B} \textbf{\bibinfo{volume}{87}},
  \bibinfo{pages}{174526} (\bibinfo{year}{2013}).

\bibitem[{\citenamefont{Renema et~al.}(2014)\citenamefont{Renema, Gaudio, Wang,
  Zhou, Gaggero, Mattioli, Leoni, Sahin, de~Dood, Fiore et~al.}}]{RenemaPRL}
\bibinfo{author}{\bibfnamefont{J.~J.} \bibnamefont{Renema}},
  \bibinfo{author}{\bibfnamefont{R.}~\bibnamefont{Gaudio}},
  \bibinfo{author}{\bibfnamefont{Q.}~\bibnamefont{Wang}},
  \bibinfo{author}{\bibfnamefont{Z.}~\bibnamefont{Zhou}},
  \bibinfo{author}{\bibfnamefont{A.}~\bibnamefont{Gaggero}},
  \bibinfo{author}{\bibfnamefont{F.}~\bibnamefont{Mattioli}},
  \bibinfo{author}{\bibfnamefont{R.}~\bibnamefont{Leoni}},
  \bibinfo{author}{\bibfnamefont{D.}~\bibnamefont{Sahin}},
  \bibinfo{author}{\bibfnamefont{M.~J.~A.} \bibnamefont{de~Dood}},
  \bibinfo{author}{\bibfnamefont{A.}~\bibnamefont{Fiore}},
  \bibnamefont{et~al.}, \bibinfo{journal}{Phys. Rev. Lett.}
  \textbf{\bibinfo{volume}{112}}, \bibinfo{pages}{117604}
  (\bibinfo{year}{2014}).

\bibitem[{\citenamefont{Zotova and Vodolazov}(2014)}]{ZotovaARXIV}
\bibinfo{author}{\bibfnamefont{A.~N.} \bibnamefont{Zotova}} \bibnamefont{and}
  \bibinfo{author}{\bibfnamefont{D.~Y.} \bibnamefont{Vodolazov}}
  (\bibinfo{year}{2014}), \eprint{arXiv:1407.3710}.

\bibitem[{\citenamefont{Vodolazov}(2014)}]{VodolazovPRB}
\bibinfo{author}{\bibfnamefont{D.~Y.} \bibnamefont{Vodolazov}},
  \bibinfo{journal}{Phys. Rev. B} \textbf{\bibinfo{volume}{90}},
  \bibinfo{pages}{054515} (\bibinfo{year}{2014}).

\bibitem[{\citenamefont{Eftekharian et~al.}(2013)\citenamefont{Eftekharian,
  Atikian, Akhlagi, Jafari~Salim, and Hamed~Majedi}}]{Eftekhariangroundstates}
\bibinfo{author}{\bibfnamefont{A.}~\bibnamefont{Eftekharian}},
  \bibinfo{author}{\bibfnamefont{H.}~\bibnamefont{Atikian}},
  \bibinfo{author}{\bibfnamefont{M.}~\bibnamefont{Akhlagi}},
  \bibinfo{author}{\bibfnamefont{A.}~\bibnamefont{Jafari~Salim}},
  \bibnamefont{and}
  \bibinfo{author}{\bibfnamefont{A.}~\bibnamefont{Hamed~Majedi}},
  \bibinfo{journal}{Appl. Phys. Lett} \textbf{\bibinfo{volume}{103}},
  \bibinfo{pages}{242601} (\bibinfo{year}{2013}).

\bibitem[{\citenamefont{Bulaevksii et~al.}(2012)\citenamefont{Bulaevksii, Graf,
  and Kogan}}]{Bulaevskii2012}
\bibinfo{author}{\bibfnamefont{L.}~\bibnamefont{Bulaevksii}},
  \bibinfo{author}{\bibfnamefont{M.}~\bibnamefont{Graf}}, \bibnamefont{and}
  \bibinfo{author}{\bibfnamefont{V.}~\bibnamefont{Kogan}},
  \bibinfo{journal}{Phys. Rev. B} \textbf{\bibinfo{volume}{85}},
  \bibinfo{pages}{014505} (\bibinfo{year}{2012}).

\bibitem[{\citenamefont{Bulaevskii et~al.}(2011)\citenamefont{Bulaevskii, Graf,
  Batista, and Kogan}}]{Bulaevskii2011}
\bibinfo{author}{\bibfnamefont{L.}~\bibnamefont{Bulaevskii}},
  \bibinfo{author}{\bibfnamefont{M.}~\bibnamefont{Graf}},
  \bibinfo{author}{\bibfnamefont{C.}~\bibnamefont{Batista}}, \bibnamefont{and}
  \bibinfo{author}{\bibfnamefont{V.}~\bibnamefont{Kogan}},
  \bibinfo{journal}{Phys. Rev. B} \textbf{\bibinfo{volume}{83}},
  \bibinfo{pages}{144526} (\bibinfo{year}{2011}).

\bibitem[{\citenamefont{Gurevich and Vinokur}(2012)}]{Gurevich2012}
\bibinfo{author}{\bibfnamefont{A.}~\bibnamefont{Gurevich}} \bibnamefont{and}
  \bibinfo{author}{\bibfnamefont{V.}~\bibnamefont{Vinokur}},
  \bibinfo{journal}{Phys. Rev. B} \textbf{\bibinfo{volume}{86}},
  \bibinfo{pages}{026501} (\bibinfo{year}{2012}).

\bibitem[{\citenamefont{Lusche et~al.}(2013{\natexlab{a}})\citenamefont{Lusche,
  Semenov, Huebers, Il'in, Siegel, Korneeva, Trifonov, Korneev, Goltsman, and
  Vodolazov}}]{Luschepreprint}
\bibinfo{author}{\bibfnamefont{R.}~\bibnamefont{Lusche}},
  \bibinfo{author}{\bibfnamefont{A.}~\bibnamefont{Semenov}},
  \bibinfo{author}{\bibfnamefont{H.}~\bibnamefont{Huebers}},
  \bibinfo{author}{\bibfnamefont{K.}~\bibnamefont{Il'in}},
  \bibinfo{author}{\bibfnamefont{M.}~\bibnamefont{Siegel}},
  \bibinfo{author}{\bibfnamefont{Y.}~\bibnamefont{Korneeva}},
  \bibinfo{author}{\bibfnamefont{A.}~\bibnamefont{Trifonov}},
  \bibinfo{author}{\bibfnamefont{A.}~\bibnamefont{Korneev}},
  \bibinfo{author}{\bibfnamefont{G.}~\bibnamefont{Goltsman}}, \bibnamefont{and}
  \bibinfo{author}{\bibfnamefont{D.}~\bibnamefont{Vodolazov}}
  (\bibinfo{year}{2013}{\natexlab{a}}), \eprint{arXiv:1303.4546}.

\bibitem[{\citenamefont{Lusche et~al.}(2013{\natexlab{b}})\citenamefont{Lusche,
  Semenov, Il'in, and Korneeva}}]{LuscheIEEE}
\bibinfo{author}{\bibfnamefont{R.}~\bibnamefont{Lusche}},
  \bibinfo{author}{\bibfnamefont{A.}~\bibnamefont{Semenov}},
  \bibinfo{author}{\bibfnamefont{K.}~\bibnamefont{Il'in}}, \bibnamefont{and}
  \bibinfo{author}{\bibfnamefont{Y.}~\bibnamefont{Korneeva}},
  \bibinfo{journal}{IEEE Trans. Appl. Supercond} \textbf{\bibinfo{volume}{23}}
  (\bibinfo{year}{2013}{\natexlab{b}}).

\bibitem[{\citenamefont{Engel et~al.}(2012)\citenamefont{Engel, Schilling,
  Il'in, and Siegel}}]{Engelmagneticfields}
\bibinfo{author}{\bibfnamefont{A.}~\bibnamefont{Engel}},
  \bibinfo{author}{\bibfnamefont{A.}~\bibnamefont{Schilling}},
  \bibinfo{author}{\bibfnamefont{K.}~\bibnamefont{Il'in}}, \bibnamefont{and}
  \bibinfo{author}{\bibfnamefont{M.}~\bibnamefont{Siegel}},
  \bibinfo{journal}{Phys. Rev. B} \textbf{\bibinfo{volume}{86}},
  \bibinfo{pages}{140506} (\bibinfo{year}{2012}).

\bibitem[{\citenamefont{Anthore et~al.}(2003)\citenamefont{Anthore, Pothier,
  and Esteve}}]{Anthore}
\bibinfo{author}{\bibfnamefont{A.}~\bibnamefont{Anthore}},
  \bibinfo{author}{\bibfnamefont{H.}~\bibnamefont{Pothier}}, \bibnamefont{and}
  \bibinfo{author}{\bibfnamefont{D.}~\bibnamefont{Esteve}},
  \bibinfo{journal}{Phys. Rev. Lett.} \textbf{\bibinfo{volume}{90}},
  \bibinfo{pages}{127001} (\bibinfo{year}{2003}).

\bibitem[{\citenamefont{Usadel}(1970)}]{Usadel}
\bibinfo{author}{\bibfnamefont{K.~D.} \bibnamefont{Usadel}},
  \bibinfo{journal}{Phys. Rev. Lett.} \textbf{\bibinfo{volume}{25}},
  \bibinfo{pages}{507} (\bibinfo{year}{1970}).

\bibitem[{\citenamefont{Bitauld et~al.}(2010)\citenamefont{Bitauld, Marsili,
  Gaggero, Mattioli, Leoni, Jahanmirinejad, L\'{e}vy, and Fiore}}]{Bitauld2010}
\bibinfo{author}{\bibfnamefont{D.}~\bibnamefont{Bitauld}},
  \bibinfo{author}{\bibfnamefont{F.}~\bibnamefont{Marsili}},
  \bibinfo{author}{\bibfnamefont{A.}~\bibnamefont{Gaggero}},
  \bibinfo{author}{\bibfnamefont{F.}~\bibnamefont{Mattioli}},
  \bibinfo{author}{\bibfnamefont{R.}~\bibnamefont{Leoni}},
  \bibinfo{author}{\bibfnamefont{S.}~\bibnamefont{Jahanmirinejad}},
  \bibinfo{author}{\bibfnamefont{F.}~\bibnamefont{L\'{e}vy}}, \bibnamefont{and}
  \bibinfo{author}{\bibfnamefont{A.}~\bibnamefont{Fiore}},
  \bibinfo{journal}{Nano Lett.} \textbf{\bibinfo{volume}{10}},
  \bibinfo{pages}{2977} (\bibinfo{year}{2010}).

\bibitem[{\citenamefont{Gaggero et~al.}(2010)\citenamefont{Gaggero,
  Jahanmirinejad, Marsili, Mattioli, Leoni, Bitauld, Sahin, Hamhuis,
  N\"{o}tzel, Sanjines et~al.}}]{Gaggero2010}
\bibinfo{author}{\bibfnamefont{A.}~\bibnamefont{Gaggero}},
  \bibinfo{author}{\bibfnamefont{S.}~\bibnamefont{Jahanmirinejad}},
  \bibinfo{author}{\bibfnamefont{F.}~\bibnamefont{Marsili}},
  \bibinfo{author}{\bibfnamefont{F.}~\bibnamefont{Mattioli}},
  \bibinfo{author}{\bibfnamefont{R.}~\bibnamefont{Leoni}},
  \bibinfo{author}{\bibfnamefont{D.}~\bibnamefont{Bitauld}},
  \bibinfo{author}{\bibfnamefont{D.}~\bibnamefont{Sahin}},
  \bibinfo{author}{\bibfnamefont{G.~J.} \bibnamefont{Hamhuis}},
  \bibinfo{author}{\bibfnamefont{R.}~\bibnamefont{N\"{o}tzel}},
  \bibinfo{author}{\bibfnamefont{R.}~\bibnamefont{Sanjines}},
  \bibnamefont{et~al.}, \bibinfo{journal}{Appl. Phys. Lett.}
  \textbf{\bibinfo{volume}{97}}, \bibinfo{pages}{151108}
  (\bibinfo{year}{2010}).

\bibitem[{Note1()}]{Note1}
Note1, \bibinfo{note}{we verified that the temperature of this custom insert
  was identical to the temperature measured on the block thermometer of the
  PPMS.}

\bibitem[{QDm()}]{QDmanual}
\emph{\bibinfo{title}{Quantum design application note 1070-207: Using ppms
  superconducting magnets at low fields}}.

\bibitem[{\citenamefont{Natarajan et~al.}(2010)\citenamefont{Natarajan,
  Peruzzo, Miki, Sasaki, Wang, Baek, Nam, Hadfield, and
  O'Brien}}]{Natarajan2010}
\bibinfo{author}{\bibfnamefont{C.~M.} \bibnamefont{Natarajan}},
  \bibinfo{author}{\bibfnamefont{A.}~\bibnamefont{Peruzzo}},
  \bibinfo{author}{\bibfnamefont{S.}~\bibnamefont{Miki}},
  \bibinfo{author}{\bibfnamefont{M.}~\bibnamefont{Sasaki}},
  \bibinfo{author}{\bibfnamefont{Z.}~\bibnamefont{Wang}},
  \bibinfo{author}{\bibfnamefont{B.}~\bibnamefont{Baek}},
  \bibinfo{author}{\bibfnamefont{S.}~\bibnamefont{Nam}},
  \bibinfo{author}{\bibfnamefont{R.~H.} \bibnamefont{Hadfield}},
  \bibnamefont{and} \bibinfo{author}{\bibfnamefont{J.~L.}
  \bibnamefont{O'Brien}}, \bibinfo{journal}{Appl. Phys. Lett.}
  \textbf{\bibinfo{volume}{96}}, \bibinfo{pages}{211101}
  (\bibinfo{year}{2010}).

\bibitem[{Note2()}]{Note2}
Note2, \bibinfo{note}{\protect \leavevmode {\protect \color {black}From the
  fact that the curves have the typical shape for 1-photon detection and from
  the low count rate, we infer that multiphoton detection events do not play a
  role. }}.

\bibitem[{\citenamefont{Lusche et~al.}(2014)\citenamefont{Lusche, Semenov,
  Korneeva, Trifonof, Korneev, Goltsman, and Hubers}}]{Luschefields}
\bibinfo{author}{\bibfnamefont{R.}~\bibnamefont{Lusche}},
  \bibinfo{author}{\bibfnamefont{A.}~\bibnamefont{Semenov}},
  \bibinfo{author}{\bibfnamefont{Y.}~\bibnamefont{Korneeva}},
  \bibinfo{author}{\bibfnamefont{A.}~\bibnamefont{Trifonof}},
  \bibinfo{author}{\bibfnamefont{A.}~\bibnamefont{Korneev}},
  \bibinfo{author}{\bibfnamefont{G.}~\bibnamefont{Goltsman}}, \bibnamefont{and}
  \bibinfo{author}{\bibfnamefont{H.}~\bibnamefont{Hubers}},
  \bibinfo{journal}{Phys. Rev. B} \textbf{\bibinfo{volume}{89}},
  \bibinfo{pages}{104513} (\bibinfo{year}{2014}).

\bibitem[{\citenamefont{Driessen et~al.}(2012)\citenamefont{Driessen, Coumou,
  Tromp, de~Visser, and Klapwijk}}]{DriessenPRL2012}
\bibinfo{author}{\bibfnamefont{E.~F.~C.} \bibnamefont{Driessen}},
  \bibinfo{author}{\bibfnamefont{P.~C. J.~J.} \bibnamefont{Coumou}},
  \bibinfo{author}{\bibfnamefont{R.~R.} \bibnamefont{Tromp}},
  \bibinfo{author}{\bibfnamefont{P.~J.} \bibnamefont{de~Visser}},
  \bibnamefont{and} \bibinfo{author}{\bibfnamefont{T.~M.}
  \bibnamefont{Klapwijk}}, \bibinfo{journal}{Phys. Rev. Lett.}
  \textbf{\bibinfo{volume}{109}}, \bibinfo{pages}{107003}
  (\bibinfo{year}{2012}).

\bibitem[{\citenamefont{Coumou et~al.}(2013)\citenamefont{Coumou, Driessen,
  Bueno, Chapelier, and Klapwijk}}]{CoumouPRB}
\bibinfo{author}{\bibfnamefont{P.~C. J.~J.} \bibnamefont{Coumou}},
  \bibinfo{author}{\bibfnamefont{E.~F.~C.} \bibnamefont{Driessen}},
  \bibinfo{author}{\bibfnamefont{J.}~\bibnamefont{Bueno}},
  \bibinfo{author}{\bibfnamefont{C.}~\bibnamefont{Chapelier}},
  \bibnamefont{and} \bibinfo{author}{\bibfnamefont{T.~M.}
  \bibnamefont{Klapwijk}}, \bibinfo{journal}{Phys. Rev. B}
  \textbf{\bibinfo{volume}{88}}, \bibinfo{pages}{180505}
  (\bibinfo{year}{2013}).

\bibitem[{\citenamefont{Bean and Livingston}(1964)}]{Beanlivingston}
\bibinfo{author}{\bibfnamefont{C.~P.} \bibnamefont{Bean}} \bibnamefont{and}
  \bibinfo{author}{\bibfnamefont{J.~D.} \bibnamefont{Livingston}},
  \bibinfo{journal}{Phys. Rev. Lett.} \textbf{\bibinfo{volume}{12}},
  \bibinfo{pages}{14} (\bibinfo{year}{1964}).

\bibitem[{\citenamefont{Plourde et~al.}(2011)\citenamefont{Plourde, van
  Harlingen, Vodolazov, Besselink, Hesselberth, and Kes}}]{Kesedgebarrier2001}
\bibinfo{author}{\bibfnamefont{B.}~\bibnamefont{Plourde}},
  \bibinfo{author}{\bibfnamefont{D.}~\bibnamefont{van Harlingen}},
  \bibinfo{author}{\bibfnamefont{D.~Y.} \bibnamefont{Vodolazov}},
  \bibinfo{author}{\bibfnamefont{R.}~\bibnamefont{Besselink}},
  \bibinfo{author}{\bibfnamefont{M.}~\bibnamefont{Hesselberth}},
  \bibnamefont{and} \bibinfo{author}{\bibfnamefont{P.~H.} \bibnamefont{Kes}},
  \bibinfo{journal}{Phys. Rev. B} \textbf{\bibinfo{volume}{64}},
  \bibinfo{pages}{014503} (\bibinfo{year}{2011}).

\bibitem[{\citenamefont{Il'in et~al.}(2010)\citenamefont{Il'in, Rall, Siegel,
  Engel, Schilling, Semenov, and Huebers}}]{Ilinwidthjc}
\bibinfo{author}{\bibfnamefont{K.}~\bibnamefont{Il'in}},
  \bibinfo{author}{\bibfnamefont{D.}~\bibnamefont{Rall}},
  \bibinfo{author}{\bibfnamefont{M.}~\bibnamefont{Siegel}},
  \bibinfo{author}{\bibfnamefont{A.}~\bibnamefont{Engel}},
  \bibinfo{author}{\bibfnamefont{A.}~\bibnamefont{Schilling}},
  \bibinfo{author}{\bibfnamefont{A.}~\bibnamefont{Semenov}}, \bibnamefont{and}
  \bibinfo{author}{\bibfnamefont{H.}~\bibnamefont{Huebers}},
  \bibinfo{journal}{Physica C} \textbf{\bibinfo{volume}{470}},
  \bibinfo{pages}{953} (\bibinfo{year}{2010}).

\bibitem[{\citenamefont{Il'in and Siegel}(2014)}]{IlinphysicaC}
\bibinfo{author}{\bibfnamefont{K.}~\bibnamefont{Il'in}} \bibnamefont{and}
  \bibinfo{author}{\bibfnamefont{M.}~\bibnamefont{Siegel}},
  \bibinfo{journal}{Physica C} \textbf{\bibinfo{volume}{503}},
  \bibinfo{pages}{58} (\bibinfo{year}{2014}).

\bibitem[{\citenamefont{Stan et~al.}(2004)\citenamefont{Stan, Field, and
  Martinis}}]{Stanmartinis}
\bibinfo{author}{\bibfnamefont{G.}~\bibnamefont{Stan}},
  \bibinfo{author}{\bibfnamefont{S.~B.} \bibnamefont{Field}}, \bibnamefont{and}
  \bibinfo{author}{\bibfnamefont{J.~M.} \bibnamefont{Martinis}},
  \bibinfo{journal}{Phys. Rev. Lett.} \textbf{\bibinfo{volume}{92}},
  \bibinfo{pages}{097003} (\bibinfo{year}{2004}).

\bibitem[{\citenamefont{Il'in et~al.}(2014)\citenamefont{Il'in, Heinrich, Luck,
  Liang, and Siegel}}]{IlinIC_PRB}
\bibinfo{author}{\bibfnamefont{K.}~\bibnamefont{Il'in}},
  \bibinfo{author}{\bibfnamefont{D.}~\bibnamefont{Heinrich}},
  \bibinfo{author}{\bibfnamefont{Y.}~\bibnamefont{Luck}},
  \bibinfo{author}{\bibfnamefont{Y.}~\bibnamefont{Liang}}, \bibnamefont{and}
  \bibinfo{author}{\bibfnamefont{M.}~\bibnamefont{Siegel}},
  \bibinfo{journal}{Phys. Rev. B} \textbf{\bibinfo{volume}{89}},
  \bibinfo{pages}{184511} (\bibinfo{year}{2014}).

\bibitem[{\citenamefont{Sac\'ep\'e et~al.}(2008)\citenamefont{Sac\'ep\'e,
  Chapelier, Baturina, Vinokur, Baklanov, and Sanquer}}]{SacepePRL}
\bibinfo{author}{\bibfnamefont{B.}~\bibnamefont{Sac\'ep\'e}},
  \bibinfo{author}{\bibfnamefont{C.}~\bibnamefont{Chapelier}},
  \bibinfo{author}{\bibfnamefont{T.~I.} \bibnamefont{Baturina}},
  \bibinfo{author}{\bibfnamefont{V.~M.} \bibnamefont{Vinokur}},
  \bibinfo{author}{\bibfnamefont{M.~R.} \bibnamefont{Baklanov}},
  \bibnamefont{and} \bibinfo{author}{\bibfnamefont{M.}~\bibnamefont{Sanquer}},
  \bibinfo{journal}{Phys. Rev. Lett.} \textbf{\bibinfo{volume}{101}},
  \bibinfo{pages}{157006} (\bibinfo{year}{2008}).

\bibitem[{\citenamefont{Noat et~al.}(2013)\citenamefont{Noat, Cherkez, Brun,
  Cren, Carbillet, Debontridder, Il'in, Siegel, Semenov, H\"ubers
  et~al.}}]{NoatPRB}
\bibinfo{author}{\bibfnamefont{Y.}~\bibnamefont{Noat}},
  \bibinfo{author}{\bibfnamefont{V.}~\bibnamefont{Cherkez}},
  \bibinfo{author}{\bibfnamefont{C.}~\bibnamefont{Brun}},
  \bibinfo{author}{\bibfnamefont{T.}~\bibnamefont{Cren}},
  \bibinfo{author}{\bibfnamefont{C.}~\bibnamefont{Carbillet}},
  \bibinfo{author}{\bibfnamefont{F.}~\bibnamefont{Debontridder}},
  \bibinfo{author}{\bibfnamefont{K.}~\bibnamefont{Il'in}},
  \bibinfo{author}{\bibfnamefont{M.}~\bibnamefont{Siegel}},
  \bibinfo{author}{\bibfnamefont{A.}~\bibnamefont{Semenov}},
  \bibinfo{author}{\bibfnamefont{H.-W.} \bibnamefont{H\"ubers}},
  \bibnamefont{et~al.}, \bibinfo{journal}{Phys. Rev. B}
  \textbf{\bibinfo{volume}{88}}, \bibinfo{pages}{014503}
  (\bibinfo{year}{2013}).

\bibitem[{\citenamefont{{Bueno} et~al.}(2014)\citenamefont{{Bueno}, {Coumou},
  {Zheng}, {de Visser}, {Klapwijk}, {Driessen}, {Doyle}, and
  {Baselmans}}}]{BuenoARXIV}
\bibinfo{author}{\bibfnamefont{J.}~\bibnamefont{{Bueno}}},
  \bibinfo{author}{\bibfnamefont{P.~C.~J.~J.} \bibnamefont{{Coumou}}},
  \bibinfo{author}{\bibfnamefont{G.}~\bibnamefont{{Zheng}}},
  \bibinfo{author}{\bibfnamefont{P.~J.} \bibnamefont{{de Visser}}},
  \bibinfo{author}{\bibfnamefont{T.~M.} \bibnamefont{{Klapwijk}}},
  \bibinfo{author}{\bibfnamefont{E.~F.~C.} \bibnamefont{{Driessen}}},
  \bibinfo{author}{\bibfnamefont{S.}~\bibnamefont{{Doyle}}}, \bibnamefont{and}
  \bibinfo{author}{\bibfnamefont{J.~J.~A.} \bibnamefont{{Baselmans}}},
  \bibinfo{journal}{ArXiv e-prints}  (\bibinfo{year}{2014}),
  \eprint{1408.0270}.

\bibitem[{\citenamefont{Engel et~al.}(2014)\citenamefont{Engel, Lonsky, Zhang,
  and Schilling}}]{EngelarXiv2}
\bibinfo{author}{\bibfnamefont{A.}~\bibnamefont{Engel}},
  \bibinfo{author}{\bibfnamefont{J.}~\bibnamefont{Lonsky}},
  \bibinfo{author}{\bibfnamefont{X.}~\bibnamefont{Zhang}}, \bibnamefont{and}
  \bibinfo{author}{\bibfnamefont{A.}~\bibnamefont{Schilling}}
  (\bibinfo{year}{2014}), \eprint{arXiv:1408.4907}.

\bibitem[{\citenamefont{Berdiyorov et~al.}(2012)\citenamefont{Berdiyorov,
  Milosevic, and Peeters}}]{Berdiyorovturns2012}
\bibinfo{author}{\bibfnamefont{G.}~\bibnamefont{Berdiyorov}},
  \bibinfo{author}{\bibfnamefont{M.}~\bibnamefont{Milosevic}},
  \bibnamefont{and} \bibinfo{author}{\bibfnamefont{F.}~\bibnamefont{Peeters}},
  \textbf{\bibinfo{volume}{100}}, \bibinfo{pages}{262603}
  (\bibinfo{year}{2012}).

\bibitem[{\citenamefont{Zotova and Vodolazov}(2013)}]{Zotovaturns2013}
\bibinfo{author}{\bibfnamefont{A.}~\bibnamefont{Zotova}} \bibnamefont{and}
  \bibinfo{author}{\bibfnamefont{D.}~\bibnamefont{Vodolazov}},
  \bibinfo{journal}{Supercond. Sci. Technol.} \textbf{\bibinfo{volume}{26}},
  \bibinfo{pages}{075008} (\bibinfo{year}{2013}).

\bibitem[{\citenamefont{Henrich et~al.}(2013)\citenamefont{Henrich, Rehm,
  Dorner, Hofherr, Il'in, Semenov, and Siegel}}]{Henrichspiral}
\bibinfo{author}{\bibfnamefont{D.}~\bibnamefont{Henrich}},
  \bibinfo{author}{\bibfnamefont{L.}~\bibnamefont{Rehm}},
  \bibinfo{author}{\bibfnamefont{S.}~\bibnamefont{Dorner}},
  \bibinfo{author}{\bibfnamefont{M.}~\bibnamefont{Hofherr}},
  \bibinfo{author}{\bibfnamefont{K.}~\bibnamefont{Il'in}},
  \bibinfo{author}{\bibfnamefont{A.}~\bibnamefont{Semenov}}, \bibnamefont{and}
  \bibinfo{author}{\bibfnamefont{M.}~\bibnamefont{Siegel}},
  \bibinfo{journal}{Applied Superconductivity, IEEE Transactions on}
  \textbf{\bibinfo{volume}{23}}, \bibinfo{pages}{2200405}
  (\bibinfo{year}{2013}).

\end{thebibliography}
\end{document}